# High-temperature QCD and the classical Boltzmann equation in curved spacetime

F. T. Brandt and J. Frenkel

*Instituto de Física, Universidade de São Paulo, São Paulo, 01498 SP, Brasil*

and

J. C. Taylor

*Department of Applied Mathematics and Theoretical Physics, University of Cambridge, Cambridge, CB3 9EW, UK*

**Abstract**

It has been shown that the high-temperature limit of perturbative thermal QCD is easily obtained from the Boltzmann transport equation for 'classical' coloured particles [2]. We generalize this treatment to curved space-time. We are thus able to construct the effective stress-energy tensor. We give a construction for an effective action. As an example of the convenience of the Boltzmann method, we derive the high-temperature 3-graviton function. We discuss the static case.

## 1  Introduction

The high-temperature behaviour of one-loop diagrams in perturbative thermal QCD is $T^2$ (where $T$ is the temperature), for any number of external gluon lines (we ignore diagrams with external quark lines in this paper) [1]. An elegant recent paper [2] has shown how the colour current $j_a^\mu(x; A)$ (where $A_\mu^a(x)$ is the gluon field) which corresponds to this infinite set of graphs may be obtained from the Boltzmann equation for classical coloured particles [3] moving in an external classical gluon field. It is also known [1,4,5] that there is a nonlocal action $W(A)$ such that

$$\sqrt{-g}\, j_a^\mu(x; A, g) = -\frac{\delta W}{\delta A_\mu^a(x)}. \tag{1.1}$$

(We have inserted a metric $g$ here anticipating step (i) below.) So far as we know, it is not obvious from the Boltzmann equation approach [2] that (1.1) should be integrable.

Here we generalize the work of [2] in the following ways:

(i) We generalize to curved spacetime, with metric $g_{\mu\nu}$. This allows us to treat graphs with just external graviton lines (with $T^4$ behaviour), and also graphs with both gluon and graviton lines ($T^2$ behaviour)

(ii) As well as the current (1.1), we are then able to define an energy tensor by

$$\sqrt{-g}T^{\mu\nu}(x;A,g) = -2\frac{\delta W}{\delta g_{\mu\nu}(x)}. \quad (1.2)$$

(iii) We write down an exact, but implicit, solution to the Boltzmann equation.

(iv) We explain why equations (1.1) and (1.2) are integrable, and obtain an expression for $W$ in terms of the stationary value of the (local) classical action for an ensemble of classical particle trajectories in the gluon and gravitational fields. This action contains both the $T^4$ and the $T^2$ terms.

(v) We discuss the positivity of $T^{00}$.

(vi) As an example of the effectiveness of the method, we derive the high-temperature 3-graviton function, which is quite complicated to obtain directly from perturbative quantum field theory.

(vii) We discuss the static case, in which explicit solutions of the Boltzmann equation are easy to write down.

## 2    The Boltzmann equation and the effective action

Following [2], we imagine an ensemble of 'classical', coloured particles, following null trajectories $x^\mu(\theta)$, where $\theta$ is an affine parameter (which we take to have dimensions of length). Each particle has a 'classical' colour charge $I^a(\theta)$ [3]. Let $p^\mu = l^{-1}\dot{x}^\mu$, where $l$ is a parameter with dimensions of length, which we use merely to keep track of dimensions. The distribution function $f(x,p,I)$ in a generalized phase-space obeys the Boltzmann equation

$$\left[ p^\mu \frac{\partial}{\partial x^\mu} - \Gamma^\lambda_{\mu\nu} p^\mu p^\nu \frac{\partial}{\partial p^\lambda} - I^a F^a_{\mu\nu}(x) g^{\nu\lambda}(x) p^\mu \frac{\partial}{\partial p^\lambda} \right.$$
$$\left. + f^{abc} I^a p^\mu A^b_\mu(x) \frac{\partial}{\partial I^c} \right] f = 0. \tag{2.1}$$

Here we have absorbed the QCD coupling constant into $A$ (in order to avoid confusion with $g = \det(g_{\mu\nu})$). We also take the dimensions of $A$ to be (length)$^{-1}$ and

$$F^a_{\mu\nu} = \partial_\mu A^a_\nu - \partial_\nu A^a_\mu + f^{abc} A^b_\mu A^c_\nu. \tag{2.2}$$

Although (2.1) is classical in the sense that it does not contain $\hbar$, it is necessary to remember that $p$ does not have the dimensions of momentum: $\hbar p$ does.

Equation (2.1) may be written in the "divergence" form

$$\left[ \frac{\partial}{\partial x^\mu} p^\mu - \frac{\partial}{\partial p^\lambda} \Gamma^\lambda_{\mu\nu} p^\mu p^\nu - \frac{\partial}{\partial p^\lambda} I^a F^a_{\mu\nu} g^{\nu\lambda} p^\mu \right.$$
$$\left. + \frac{\partial}{\partial I^c} f^{abc} I^a p^\mu A^b_\mu \right] g(x) f = 0. \tag{2.3}$$

This is a sort of Liouville theorem, and shows that

$$d^4x \, d^4p \, dI \, g(x) \tag{2.4}$$

is an invariant phase-space element along a trajectory (for the definition of $dI$ see [2]). The current and energy tensor are then given by

$$j^{a\mu}(x) = \int d^4p \, dI \sqrt{-g} \, p^\mu I^a f(x, p, I), \tag{2.5}$$

$$T^{\mu\nu}(x) = \int d^4p \, dI \sqrt{-g} \, p^\mu p^\nu f(x, p, I). \tag{2.6}$$

From (2.1) follow the conservation equations

$$D_\nu j^{a\nu} + f^{abc} A^b_\nu j^{c\nu} = 0, \tag{2.7}$$

$$D_\nu T^{\mu\nu} + F^{\mu\nu}_a g_{\nu\lambda} j^{a\lambda} = 0, \tag{2.8}$$

where $D$ denotes the coordinate (not colour) covariant derivative.

Consider now a family of trajectories $(x(\theta), p(\theta), I(\theta))$, and define $(Y, Q, I_0)$ by the conditions

$$x \sim Y, \quad p \to l^{-1}Q, \quad I \to I_0 \quad (\text{as } x \to \infty) \tag{2.9}$$

and

$$\dot{x}^\mu = Q^\alpha \frac{\partial x^\mu}{\partial Y^\alpha}. \tag{2.10}$$

Thus three of the components of $Y$ parametrize the position of the trajectory at infinity, and the fourth replaces $\theta$ as an affine parameter along a trajectory. $Y$ is a vector in the asymptotic Minkowski space (we assume that the colour and gravitational fields tend to zero at infinity). Then there are relations

$$x = x(Y, Q, I_0), \quad p = p(Y, Q, I_0), \quad I = I(Y, Q, I_0). \tag{2.11}$$

Any function of a constant of the motion satisfies the Boltzmann equation (2.1), and an example of such a constant is $Q^\alpha$ itself. We seek a solution of (2.1) which reduces to the equilibrium distribution at temperature $T$ at infinity. The required solution is evidently

$$f(x, p, I) = \frac{C\hbar^4}{\hbar^3} N[Q(x, p, I)] \, 2 \, l^2 \, \delta(Q^2) \, \theta(Q_0), \tag{2.12}$$

where

$$N(Q) = \frac{1}{\exp(\hbar Q_0/lT) \pm 1} \tag{2.13}$$

and $C$ gives the number of spin and internal degrees of freedom. Here the factor $\hbar^4$ is because of the $d^4p$ in (2.4), and the factor $l^2$ is because of the $l^{-1}$ in (2.9). Equation (2.12) is the exact solution of (2.1) with which we will work.

We may now write equations for $j^\mu$ and $T^{\mu\nu}$:

$$\begin{aligned}
d^4x\sqrt{-g}j^{a\mu}(x)\delta A^a_\mu(x) &= d^4x \int d^4p \, dI \, g(x) f(x, p, I) \, p^\mu \, I^a \, \delta A^a_\mu(x) \\
&= \frac{2C\hbar}{(2\pi)^3 l^2} d^4Y \int d^4Q dI_0 N(Q) \delta(Q^2) \theta(Q^0) \\
&\quad \times p^\mu(Y, Q, I_0) I^a(Y, Q, I_0) \delta A^a_\mu(x(Y, Q, I_0)).
\end{aligned} \tag{2.14}$$

So that

$$\sqrt{-g}j^{a\mu}(x) = \frac{2C\hbar}{l^2(2\pi)^3} \int d^4Q dI_0 N(Q)\delta(Q^2)\theta(Q^0)\Delta(Y(x,Q,I_0),Q,I_0)$$
$$\times p^\mu(Y(x,Q,I_0),Q,I_0)I_0^a(Y(x,Q,I_0),Q,I_0), \quad (2.15)$$

where

$$\Delta(Y,Q,I_0) = \left[\det\left(\frac{\partial x^\mu(Y,Q,I_0)}{\partial Y^\alpha}\right)\right]^{-1}. \quad (2.16)$$

Similarly

$$\sqrt{-g}T^{\mu\nu}(x) = \frac{2C\hbar}{l^2(2\pi)^3} \int d^4Q dI_0 N(Q)\delta(Q^2)\theta(Q^0)\Delta(Y(x,Q,I_0),Q,I_0)$$
$$\times p^\mu p^\nu(Y(x,Q,I_0),Q,I_0). \quad (2.17)$$

In order to evaluate these expressions in perturbation theory, one needs to find the functions in (2.12) in perturbation theory, and deduce perturbation expansions for $Y(x,Q,I_0)$ and $\Delta$. In contrast to the method of expansion used in [2], no derivatives of $N(Q)$ appear in our method. (Of course, the two expansions must be equivalent when one uses the $Q$-integration.)

We now construct an effective action $W(A,g)$ such that (2.15) and (2.17) are given by (1.1) and (1.2). There is no action for Wong's equations [3] in terms of $I^a$ (see for example [6] and references therein). Instead we introduce a quantity $u$, transforming say as the (complex) $n$-dimensional fundamental representation of SU($n$) (which we now assume the colour group to be), with representation matrices $T^a$. This is related to $I$ by

$$I^a = u^* T^a u. \quad (2.18)$$

We consider $u$ to be a dynamical variable along the trajectory, with asymptotic value $u_0$. Then we take the action

$$\tilde{S}(Q,u_0; x, u; A, g) =$$
$$\frac{\hbar}{l^4}\int d^4Y \left[\{\dot{x}^\mu Q_\mu + ilu_0^*\dot{u}\} - \frac{1}{2}g_{\mu\nu}(x)\dot{x}^\mu\dot{x}^\nu - ilu^*\dot{u} - lu^*T^a u \dot{x}^\mu A_\mu^a(x)\right], \quad (2.19)$$

where the 'dot' differentiation was defined in (2.10). Thus $x(Y)$ and $u(Y)$ appear in (2.19) like 'fields' in $Y$-space. The term in the brace in (2.19), being a differential, does not affect the equations of motion. The reason for introducing it will be explained later. In (2.19), factors of $l$ are introduced to make everything dimensionless, and the factor $\hbar$ to give the whole thing the dimensions of an action.

The equations of motion got by varying $x(Y)$ and $u(Y)$ in (2.19) are

$$i\dot{u} + \dot{x}^\mu A_\mu^a(x)T^a u = 0, \tag{2.20}$$

$$g_{\mu\lambda}(x)[\ddot{x}^\lambda + \Gamma_{\nu\sigma}^\lambda(x)\dot{x}^\nu \dot{x}^\sigma] + lI^a\left(\frac{\partial A_\mu^a}{\partial x^\nu} - \frac{\partial A_\nu^a}{\partial x^\mu}\right)\dot{x}^\nu - l\dot{I}^a A_\mu^a(x) = 0, \tag{2.21}$$

where we have used (2.18). From (2.19) we deduce the Wong equation [3]

$$\dot{I}^a = f^{abc}I^b A_\nu^c(x)\dot{x}^\nu, \tag{2.22}$$

and using this Eq. (2.21) gives the expected equation of motion.

The next step is to define the effective action

$$S(Q, u_0; A, g) = \tilde{S}[Q, u_0; \xi(Q, u_0; A, g), \upsilon(Q, u_o; A, g); A, g], \tag{2.23}$$

where $x = \xi$ and $u = \upsilon$ are solutions of the equations of motion (2.20) and (2.21) subject to the boundary conditions (2.9) and $u \to u_0$. Now make infinitesimal variations in $A$ and $g$. We have

$$\begin{aligned}S(A + \delta A, g + \delta g) &= \tilde{S}[\xi + \delta\xi, \upsilon + \delta\upsilon; A + \delta A, g + \delta g] \\ &= \tilde{S}[\xi, \upsilon; A + \delta A, g + \delta g],\end{aligned} \tag{2.24}$$

where

$$\xi + \delta\xi = \xi(A + \delta A, g + \delta g), \quad \upsilon + \delta\upsilon = \upsilon(A + \delta A, g + \delta g), \tag{2.25}$$

and in the second equality in (2.24) we have made use of the fact that $x = \xi$, $u = \upsilon$ is a stationary value of $\tilde{S}$. Thus we deduce from (2.19) and (2.24) that

$$\begin{aligned}\frac{\hbar}{l^2}\int d^4Y \delta A_\mu^a(x(Y))I^a p^\mu &= \tilde{S}(A + \delta A, g) - \tilde{S}(A, g) \\ &= S(A + \delta A, g) - S(A, g).\end{aligned} \tag{2.26}$$

Comparing this with (2.14), we see that (1.1) is satisfied by

$$W = \frac{2C}{(2\pi)^3}\int d^4Q du_0 \delta(Q^2)\theta(Q_0)N(Q)S(Q, u_0; A, g). \tag{2.27}$$

An exactly similar argument shows that (2.27) also satisfies (1.2).

We now explain the reason for inserting the brace into (2.19). The proof of the second equality in (2.24) involves an integration by parts, in which we require that

$$\int d^4 Y Q \cdot \frac{\partial}{\partial Y}[\{g_{\mu\nu}\dot{\xi}^\mu - Q_\mu\}\delta\xi + li\{v^* - v_0^*\}\delta v] = 0. \tag{2.28}$$

This would not be true without the contributions from the brace in (2.19), since $\delta\xi$ and $\delta v$ do not tend to zero fast enough for large $Y$. In fact the *whole* contribution to (2.27) comes from the brace in (2.19). The other terms in (2.19) are zero for a solution of (2.20) and for a null-trajectory solution of (2.21).

In order to use (2.27) to get the effective actions of hard-thermal loops, we need to expand (2.23) in powers of $l$. The leading term is of order $l^{-4}$. To this order, $W$ is independent of the gluon field $A$. The $O(1)$ term in (2.21) has to be solved exactly to find this contribution to $W$. The effective action is that for hot matter in curved spacetime. It is, from (2.13), of order $\hbar(T/\hbar)^4$. The next nonvanishing terms in (2.23) are order $l^{-2}$. To find them, we may solve (2.21) iteratively in powers of $l$, but (2.22) must be solved to all orders. The contribution from (2.27) is order $\hbar(T/\hbar)^2$. This is the effective action for gluons in curved spacetime. One might continue the expansion in powers of $l$, but for massless bosons the lower order terms may be infrared divergent.

Finally, note that $T^{00}$ from (2.17) is positive. It follows that the leading, $l^{-4}$ term is positive. But we cannot say anything about the positivity of the $l^{-2}$ term on its own.

## 3 The 3-graviton hard thermal loop

In practice, iterative solution of (2.1) seems to be an efficient way of calculating hard thermal contributions. As an example, we derive the 3-graviton function (keeping just the graviton terms in (2.1)). We have found it convenient to use the variable $p_\mu = g_{\mu\nu}(x)p^\nu$, instead of $p^\mu$. We define

$$F(x^\mu, p_\nu, I^a) = f(x^\mu, g_{\lambda\nu}(x)p^\lambda, I^a) \tag{3.1}$$

Then the factor $g(x)$ is omitted from (2.4), and the relevant part of (2.1) becomes

$$p^\alpha \left(\partial_\alpha + \Gamma^\gamma_{\alpha\beta} p_\gamma \frac{\partial}{\partial p_\beta}\right) F(x, p, I) = 0. \tag{3.2}$$

Using the well-known expression of the Christofell symbol $\Gamma^{\gamma}_{\alpha\beta}$, and defining the graviton field

$$\phi^{\alpha\beta} \equiv g^{\alpha\beta} - \eta^{\alpha\beta}, \tag{3.3}$$

the Boltzmann equation can be written in the form

$$p \cdot \partial \, F(x,p,I) = \hat{L} F(x,p,I) \tag{3.4}$$

where $p \cdot \partial \equiv \eta^{\mu\nu} p_\mu p_\nu$ and $\hat{L}$ is the differential operator

$$\hat{L} \equiv \frac{1}{2}\left[\left(\partial_\gamma \phi^{\alpha\beta}\right) p_\alpha p_\beta \frac{\partial}{\partial p_\gamma} - \phi^{\alpha\beta}\left(p_\alpha \partial_\beta + p_\beta \partial_\alpha\right)\right], \tag{3.5}$$

which is *linear* in the graviton field $\phi^{\alpha\beta}$.

It is now easy to find recursively the solution of the Boltzmann equation, which is given by

$$F^{(0)} = 2C\hbar\theta(p_0)\,\delta(\eta^{\mu\nu} p_\mu p_\nu)\,N(p_0)$$
$$F^{(n)} = \left(\frac{1}{p\cdot\partial}\hat{L}\right)^n F^{(0)}, \tag{3.6}$$

where $F^{(n)}$ is of order $\phi^n$.

The $n$-point graviton functions can be obtained by functional differentiation of the generating action using the relation

$$\frac{\delta W}{\delta g^{\mu\nu}} = \frac{1}{2}\sqrt{-g}T_{\mu\nu} = \frac{1}{2}\int \frac{\mathrm{d}^4 p}{(2\pi)^3} p_\mu p_\nu F(x,p). \tag{3.7}$$

From Eqs. (3.5) and (3.6) we can see that *all* $F^{(n)}$ have degree $(-2)$ in $p$. Making the rescaling $p \to p/|\mathbf{p}| \equiv Q$, the n-graviton functions, in the momenta space, can be written in terms of dimensionless functions $\tilde{f}_{\mu_1\nu_1\cdots\mu_n\nu_n}(k^2,\cdots,k^n,Q)$ as

$$\Pi_{\mu_1\nu_1\cdots\mu_n\nu_n}\left(k^1,\cdots,k^n\right) = \rho\int \frac{\mathrm{d}}{\Omega} 4\pi \frac{(n-1)!}{2}\delta\left(k^1+\cdots+k^n\right)$$
$$\times \tilde{f}_{\mu_1\nu_1\cdots\mu_n\nu_n}\left(k^2,\cdots,k^n,Q\right)\Big|_{Q^2=0}, \tag{3.8}$$

where the integration is over the directions of the 3-vector $\mathbf{Q}$ and

$$\rho = \frac{C\hbar}{2\pi^2} \int_0^\infty \frac{|\mathbf{p}|^3 \, \mathrm{d}\,|\mathbf{p}|}{\exp(|\hbar\mathbf{p}|/T) - 1} = C\frac{\pi^2 T^4}{30\hbar^3} \tag{3.9}$$

is the energy density of the thermal particles.

The functions $\tilde{f}_{\mu_1\nu_1\cdots\mu_n\nu_n}(k^2, \cdots, k^n, Q)$ denote the coefficient of $\tilde{\phi}^{\mu_2\nu_2}(k^2)\cdots\tilde{\phi}^{\mu_n\nu_n}(k^n)$ in the integrand of (3.7), and can be obtained in a systematic way solving Eq. (3.6) for $n = 1, \ 2, \ \cdots$. Let us first consider the two-graviton function. In momentum space, Eq. (3.6) gives

$$\begin{aligned}
p_\mu p_\nu F^{(1)}(k,p) &= \phi^{\alpha\beta}(k) \tilde{f}_{\mu\nu\alpha\beta}(k,p) = \frac{\tilde{\phi}^{\alpha\beta}(k)}{2} \frac{p_\mu p_\nu p_\alpha p_\beta}{p \cdot k} k_\gamma \frac{\partial F^{(0)}}{\partial p_\gamma} \\
&= \frac{\tilde{\phi}^{\alpha\beta}(k)}{2} \left[ k \cdot \partial_p \left( \frac{p_\mu p_\nu p_\alpha p_\beta}{p \cdot k} F^{(0)} \right) - F^{(0)} k \cdot \partial_p \frac{p_\mu p_\nu p_\alpha p_\beta}{p \cdot k} \right],
\end{aligned} \tag{3.10}$$

where $k \cdot \partial_p \equiv k_\alpha \partial/\partial p_\alpha$. Using the on-shell constraint, we can see that the surface term in the last equation can be neglected because in the limit $p_0 \to \infty$ the function $N(p_0)$ in the first equation of (3.6) goes much faster to zero than $p_0^3 = |\mathbf{p}|^3$. This will also be true for all $n$-graviton functions. From Eq. (3.10) one can readily obtain $\tilde{f}_{\mu\nu\alpha\beta}(k, \ Q)$. Then, Eq. (3.8) yields the following result for the 2-graviton function

$$\Pi_{\mu\nu\alpha\beta}(k) = -\rho \int \frac{\mathrm{d}\Omega}{4\pi} \left( \frac{1}{4} k \cdot \partial_Q \frac{Q_\mu Q_\nu Q_\alpha Q_\beta}{Q \cdot k} \right)\bigg|_{Q^2=0}. \tag{3.11}$$

This result is in agreement with the hard thermal loop 2-graviton function obtained by standard Feynman diagrammatic calculation [7]. The term inside the round bracket is identical to the *forward scattering amplitude* of a *hard* on-shell thermal particle off two external graviton fields.

Let us now consider the 3-graviton function. Using Eq. (3.6) with $n = 2$ and adding total derivatives we obtain in the momentum space

$$\begin{aligned}
p_\mu p_\nu F^{(2)} &= \tilde{\phi}^{\alpha\beta}\left(k^2\right) \tilde{\phi}^{\rho\sigma}\left(k^3\right) \tilde{f}_{\mu\nu\alpha\beta\rho\sigma}\left(k^2,\ k^3\ p\right) \\
&= \frac{\tilde{\phi}^{\alpha\beta}(k^2) \tilde{\phi}^{\rho\sigma}(k^3)}{4} \Bigg\{ F^{(0)} k^3 \cdot \partial_p \left[ \frac{p_\rho p_\sigma}{p \cdot k^3} k^2 \cdot \partial_p \frac{p_\mu p_\nu p_\alpha p_\beta}{p \cdot (k^2 + k^3)} \right. \\
&\quad \left. + \frac{p_\mu p_\nu p_\rho p_\sigma \left(p_\alpha k_\beta^3 + p_\beta k_\alpha^3\right)}{p \cdot k^3 p \cdot (k^2 + k^3)} \right] \Bigg\},
\end{aligned} \tag{3.12}$$

Making $p \to Q$ in the coefficient of $\tilde{\phi}^{\alpha\beta}(k^2)\tilde{\phi}^{\rho\sigma}(k^3)$ and using Eq. (3.8), we obtain

$$\Pi_{\mu\nu\alpha\beta\rho\sigma}\left(k^2,\ k^3\right) = \rho \int \frac{d\Omega}{4\pi} \mathcal{A}_{\mu\nu\alpha\beta\rho\sigma}\left(k^2,\ k^3,\ Q\right), \tag{3.13}$$

where

$$\mathcal{A}_{\mu\nu\alpha\beta\rho\sigma}\left(k^2,\ k^3,\ Q\right) = \frac{1}{8}k^3 \cdot \partial_Q \left( \frac{Q_\rho Q_\sigma}{Q \cdot k^3} k^2 \cdot \partial_Q \frac{Q_\mu Q_\nu Q_\alpha Q_\beta}{Q \cdot (k^2 + k^3)} \right.$$
$$\left. + \frac{Q_\mu Q_\nu Q_\rho Q_\sigma \left(Q_\alpha k^3_\beta + Q_\beta k^3_\alpha\right)}{Q \cdot k^3\ Q \cdot (k^2 + k^3)} \right) \bigg|_{Q^2=0}$$
$$+ \left(k^2,\ \alpha,\ \beta\right) \longleftrightarrow \left(k^3,\ \rho,\ \sigma\right). \tag{3.14}$$

A lengthy calculation (we have used computer algebra) has shown explicitly that (3.14) is the same as what is obtained from high-temperature perturbative quantum field theory and is therefore symmetric under all permutations of

$$(k^1, \mu, \nu),\ \ (k^2, \alpha, \beta),\ \ (k^3, \rho, \sigma). \tag{3.15}$$

This verification makes use of eikonal identities like

$$\frac{1}{Q \cdot k^1 Q \cdot k^3} + \frac{1}{Q \cdot k^1 Q \cdot k^2} + \frac{1}{Q \cdot k^2 Q \cdot k^3} = 0. \tag{3.16}$$

We stress that it would be very hard to guess, directly from perturbative quantum field theory that the 3-graviton function could be written in such a simple form as given by (3.14) (compare with the second work in reference [7]).

## 4 The static case

The Boltzmann equation (2.1) is of course simple to solve in the static case. It is easy to verify that

$$p_0 + I^a A_0^a(\mathbf{x}) \tag{4.1}$$

is a constant of the motion. For reasons which will become clear, we now allow for a nonzero mass $m$. Then, in the notation of (3.1), an appropriate solution of the Boltzmann equation is

$$F = \frac{2C\hbar}{(2\pi)^3}\delta(g^{\mu\nu}p_\mu p_\nu - m^2/\hbar^2))\theta(p_0)N\left[l\left(p_0 + I^a A_0^a(\mathbf{x})\right)\right], \tag{4.2}$$

where $N$ is defined in (2.13). For the Bose case, in order to ensure that $F$ is positive, we must restrict $A$ to satisfy

$$|I^a A_0^a(\mathbf{x})| < m/\hbar. \tag{4.3}$$

With (4.2), equations (2.5) and (2.6) may be used to define $j^\mu$ and $T_{\mu\nu}$. Using the change of variable (4.5) below, it follows that $j^i = 0$. With this information, one may verify that that equations (1.1) and (1.2) may be integrated to give

$$W = \frac{C\hbar}{(2\pi)^3}\int d^4x \int d^4p \int dI\, \theta(g^{\mu\nu}(x)p_\mu p_\nu - m^2/\hbar^2)$$
$$\times \theta(p_0)N\left[l\left(p_0 + I^a A_0^a(\mathbf{x})\right)\right]. \tag{4.4}$$

The **p**-integration may be done by the substitution

$$p'_i = p_i - g_{i0}p_0/g_{00}, \tag{4.5}$$

which gives

$$g^{\mu\nu}p_\mu p_\nu = (g_{00})^{-1}p_0^2 - g^{ij}p'_i p'_j, \tag{4.6}$$

and thus

$$W = \frac{C\hbar}{6\pi^2}\int d^4x \frac{\sqrt{-g}}{g_{00}^2}\int dI \int_r^\infty dp_0 [p_0^2 - r^2]^{3/2} N\left[l\left(p_0 + I^a A_0^a(\mathbf{x})\right)\right], \tag{4.7}$$

where $r = m\sqrt{g_{00}}/\hbar$.

After expanding $N$ as a power series in exponentials, the $p_0$-integration may be done, to obtain

$$W = \frac{C\hbar}{\pi^2}\left(\frac{T}{\hbar}\right)^4 \int d^4x \frac{\sqrt{-g}}{g_{00}^2}(\pm)\int dI$$
$$\times \sum_{n=1}^{\infty} L_n(m\sqrt{g_{00}}/T)n^{-4}[\pm\exp(-\hbar I^a A_0^a/T)]^n, \tag{4.8}$$

where

$$L_n = \frac{1}{2} (nm\sqrt{g_{00}}/T)^2 \, K_2(nm\sqrt{g_{00}}/T) \tag{4.9}$$

$K_2$ being the modified Bessel function [8]. These coefficients have the properties

$$L_n \to 1 \quad \text{for} \quad n \ll \frac{T}{m\sqrt{g_{00}}},$$
$$L_n \sim \exp(-nm\sqrt{g_{00}}/T) \quad \text{for} \quad n \gg \frac{T}{m\sqrt{g_{00}}}. \tag{4.10}$$

We now consider the Fermi and Bose cases separately. For the Fermi case, there is no difficulty in going to the limit $m = 0$. Then the gluon dependence of (4.8) may be expressed in terms of the generalized Riemann zeta function $\Phi$ [8]

$$-\int dI \sum_{n=1}^{\infty} n^{-4} \left[-\exp(-\hbar I^a A_0^a/T)\right]^n =$$
$$\int dI \exp(-\hbar I^a A_0^a/T) \, \Phi\left[-\exp(-\hbar I^a A_0^a/T), 4, 1\right]. \tag{4.11}$$

The integration over $I$ in (4.11) depends upon the colour group [2]. For the case of SU(2), this integration may be made explicit. Writing $A_0^a I^a = A_0 I x$, we may use

$$\int dI \cdots = \Lambda_{B,F} \int_{-1}^{1} dx \cdots, \tag{4.12}$$

with $\Lambda_B = 3/2$ and $I = \sqrt{2}$ for gluons and $\Lambda_F = 1$ and $I = \sqrt{3}/2$ for quarks. In this case there are only even powers of the gluon field in (4.11). The $A_0^2$ term is the familiar $T^2$ electric gluon mass term.

One may ask under what conditions the result (4.11) is expected to be relevant (aside from the static condition). Since no derivatives of the gluon field appear, one expects to need

$$k \ll |I^a A_0^a|, \quad k \ll T/\hbar, \tag{4.13}$$

where $k$ is a typical gluon wave-number. Also, in order that higher powers of $A_0$ than the second should be relevant, we need

$$|I^a A_0^a| \sim T/\hbar. \tag{4.14}$$

Now turn to the Bose case, with condition (4.3). Expanding the exponentials in (4.8), we obtain a series of the form

$$W = C\hbar \int d^4 x \frac{\sqrt{-g}}{g_{00}^2} \int dI \sum_{n=0}^{\infty} c_n \left(T, m\sqrt{g_{00}}\right) \left(I^a A_0^a\right)^n, \tag{4.15}$$

where, for $m \ll T$,

$$c_0 \to \frac{\pi^2}{90} \left(\frac{T}{\hbar}\right)^4, \qquad c_2 \to \frac{1}{12} \left(\frac{T}{\hbar}\right)^2,$$

$$c_3 \to a_3 \left(\frac{T}{\hbar}\right) \log\left(m\sqrt{g_{00}}/T\right), \quad c_n \to a_n \left(\frac{T}{\hbar}\right) \left(m\sqrt{g_{00}}/\hbar\right)^{3-n} \quad (n > 3), \tag{4.16}$$

where $a_3, a_4, \cdots$ are numerical constants. For instance, we find that $a_3 \simeq 0.2$ and $a_4 \simeq 0.1$.

The first term in (4.15) corresponds to the action for hot mater in curved space-time. The linear term in the gauge field yields a vanishing contribution when the integration over colour is performed. The second order term in $A_0$ is a good approximation if

$$|I^a A_0^a(\mathbf{x})| < m/\hbar \ll T/\hbar. \tag{4.17}$$

Higher order contributions involving the gluon fields become significant when the terms in (4.17) are of comparable orders of magnitude (and (4.13) holds).

**Acknowledgement**

JCT acknowledges helpful advice from Roman Jackiw. This research is supported in part by the EU Programme "Human Capital and Mobility", Network "Physics at High Energy Colliders", contract CHRX-CT93-0357 (DG12 COMA), and by PPARC and CNPq.